\documentstyle[epsfig]{article}

\newcommand{\fig}[9]{
\begin{figure} [htpb]
\begin{center}
\mbox{\epsfig{figure=#3,height=#4 mm,%
width=#5 mm,bbllx=#6 pt,bblly=#7 pt,bburx=#8 pt,bbury=#9 pt}}
\caption{#1}
\label{#2}
\end{center}
\end{figure}}

\title{CANDELA PHOTO-INJECTOR EXPERIMENTAL RESULTS}

\author{C. Travier, L. Boy\thanks{Present address: GANIL, BP 5027,
F-14021 Caen}, J.N. Cayla, B. Leblond\\
        Laboratoire de l'Acc\'el\'erateur Lin\'eaire\\
        IN2P3-CNRS et Universit\'e de Paris-Sud, B\^at. 200\\
	F-91405 Orsay\\
and\\
        P. Georges, P. Thomas\\
        Institut d'Optique Th\'eorique et Appliqu\'ee\\
        Universit\'e de Paris-Sud, B\^at. 501\\
F-91405 Orsay}

\begin{document}

\maketitle                   

\begin{abstract}
The CANDELA photo-injector is a two cell S-band photo-injector. The copper
cathode is illuminated by a 500 fs Ti:sapphire laser. This paper presents
energy spectrum measurements of the dark current and intense electron emission
that occurs when the laser power density is very high.
\end{abstract}

\vspace{0.4cm}
\centerline{\large\bf Introduction}
\vspace{0.4cm}
The CANDELA photo-injector is an RF gun made of two decoupled 3 GHz cells
\cite{bib:lal,bib:lal1,bib:lal2}. The
Ti:sapphire laser system used to illuminate the photocathode was
designed by the "Institut d'Optique Th\'eorique et Appliqu\'ee"
at Orsay \cite{bib:lal3}. CANDELA was first operated at the end of 1993,
thus being the "first
femtosecond laser driven photo-injector"\cite{bib:lallondres}.
A maximum charge of 0.11 nC was extracted corresponding to an effective
quantum efficiency of 5 $\times 10^{-6}$.
Since these experiments, we have had several problems
with the laser system, so that we could not do many new experiments.
The results presented in this paper are divided in two parts:
new measurements on dark current, especially the energy
spectrum and new measurements on photo-electrons, showing that above
a certain laser power density, intense electron emission is observed.

\vspace{0.4cm}
\centerline{\large\bf Experimental setup}
\vspace{0.4cm}

The gun is powered by an old
THOMSON TV2013 klystron that can deliver a maximum measured peak power of
2.6 MW. This power can be shared between the two cells with an arbitrary
ratio. The RF phase
between the two cells can be freely adjusted by a high power phase
shifter.

As shown in figure \ref{figexp}, the gun is followed by a short beamline
including the different
diagnostics used to measure the parameters of the beam
(charge, position, transverse density, emittance, energy, energy spread
and bunch length). Not all the diagnostics were completed at the
time of the experiments. Only the following ones
were available:
2 wall current monitors (WCM) to measure the charge and position
of photo-electron pulses,
a coaxial Faraday cup for charge measurement on the straight
line behind the first dipole,
a slit with isolated jaws to measure the energy, and a
fluorescent screen with a TV camera for beam transverse
observation.

\fig{Experimental setup}{figexp}{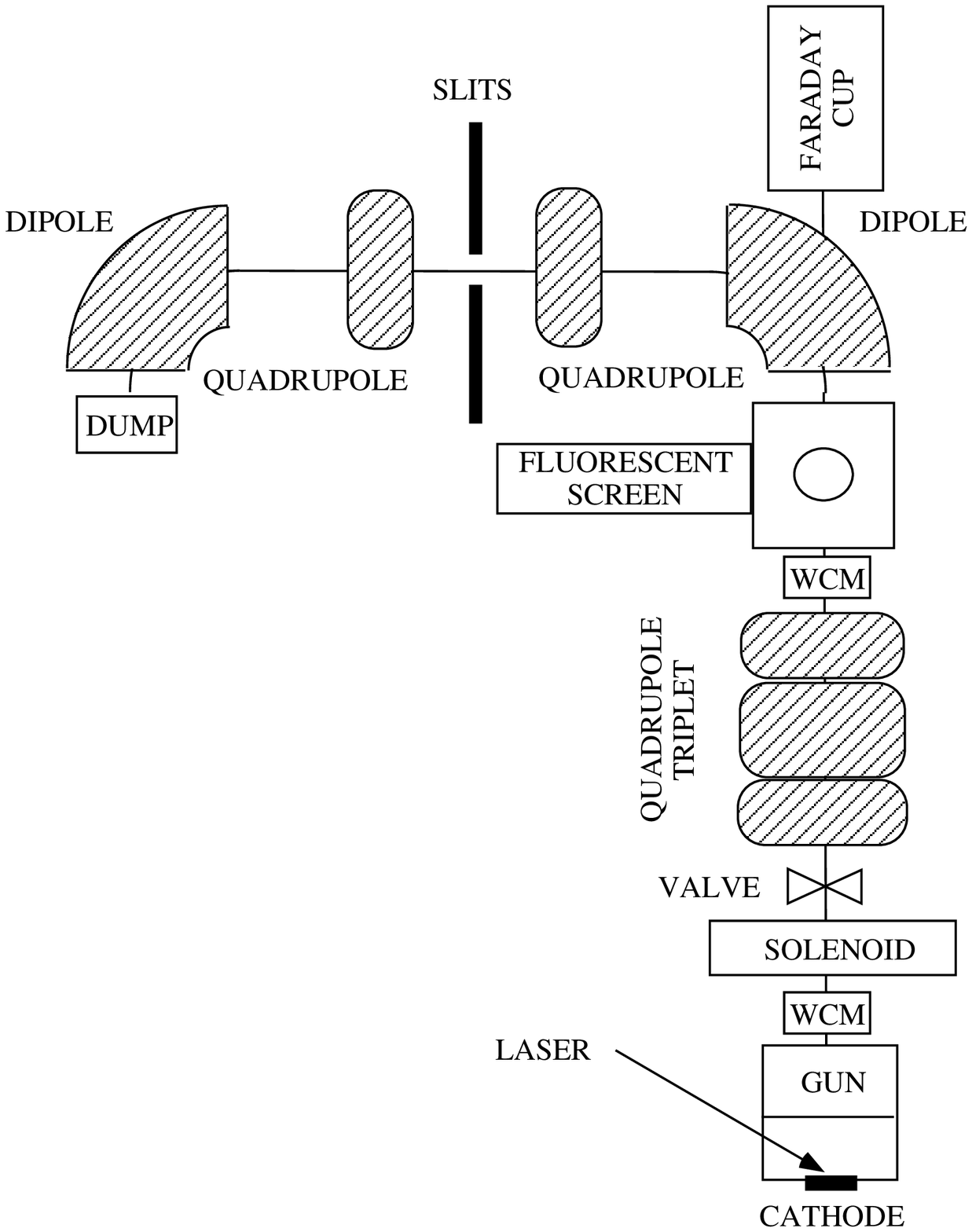}
{112.1}{90}{90.72}{155.925}{544.32}{720.09}

The laser is a Ti:sapphire laser described in reference \cite{bib:lal3}.
It produces one single pulse (at 12.5 Hz), with an adjustable
duration from 150 fs to 15 ps, with a
maximum energy of 2.5 mJ at 800 nm.  Figure \ref{figauto}
presents an autocorrelation trace of a 280 fs pulse produced at 800 nm.
225 $\mu$J of UV light (260 nm) is then
obtained through third harmonic generation in BBO crystals. Though no pulse
length measurements
can be done on UV pulses, one can assume that due
to dispersion in the crystals, the laser is somewhat longer
(typically 500 fs).
The optical path between the laser
room and the gun cathode (around 25 m long) includes several lenses
and mirrors.
The overall measured efficiency of the light transport system is
75\%.
The laser light is injected into the gun via one of the two 54$^\circ$30'
entry ports. The laser is focused onto the cathode and different
spot sizes can be obtained by changing the last lens position. No UV camera
was available to measure the exact spot size.

\fig{Autocorrelation trace of the IR laser pulse}
{figauto}{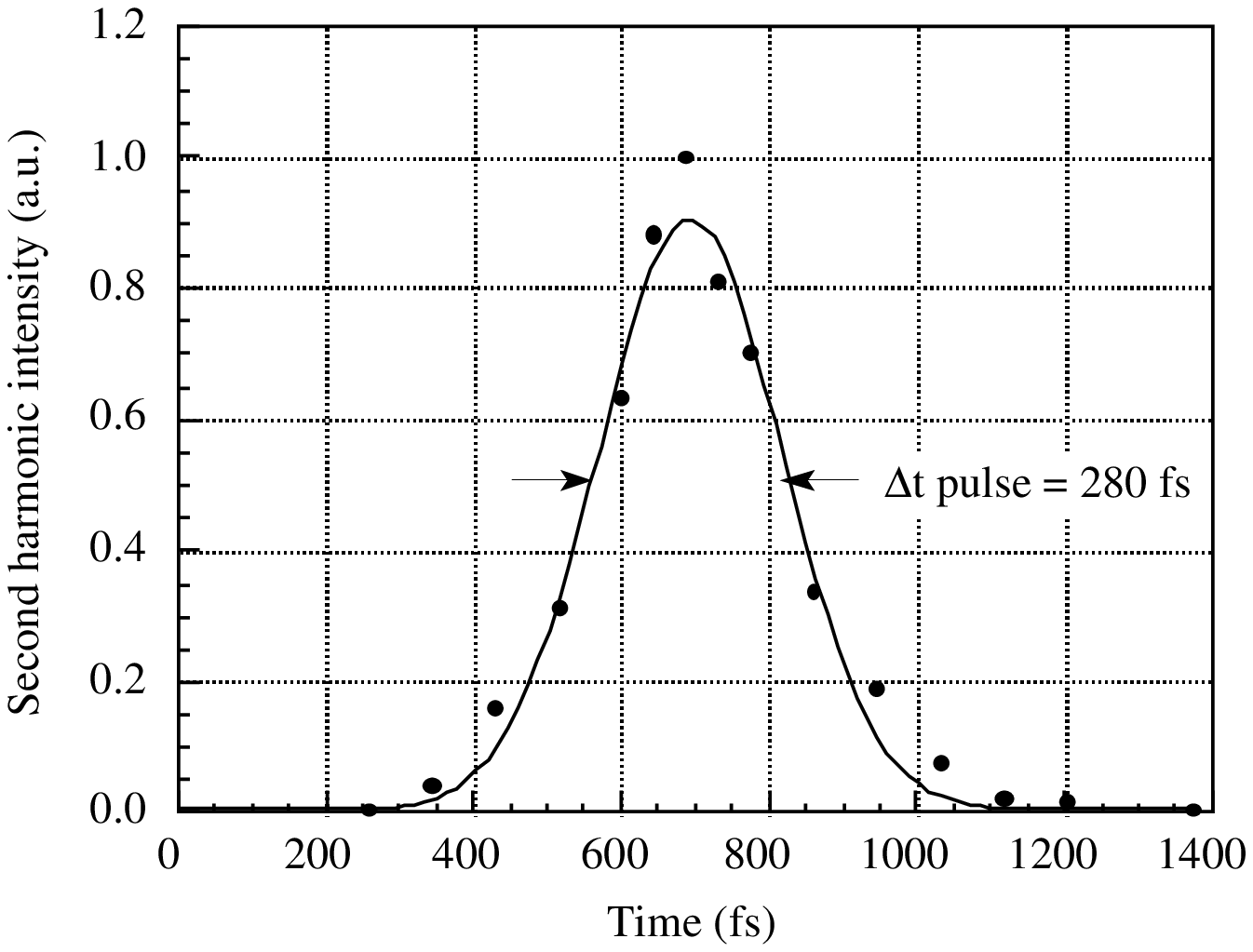}{68.45}{90}{119.07}{419.58}{524.475}{728.595}

Normally, the laser is synchronized to the RF frequency via an
electronic feedback
loop that is driving a piezo-electric transducer that adjusts the laser
oscillator cavity length. However, at the time of experiments reported
here, this system was not working, so that the relative phase between the
laser and the RF was random.

\vspace{0.4cm}
\centerline{\large\bf Dark current measurements}
\vspace{0.4cm}

When an RF cavity is submitted to very high fields, electrons are emitted
through the field emission process. Some of them are captured and
accelerated, thus resulting in a parasitic current named dark current.
Results on the dark current measurement of the CANDELA RF gun have
already been published \cite{bib:lal4}. However, at the time of these
first measurements, it was not possible to measure the energy spectrum
of this dark current. These spectra are now obtained
by reading the current flowing through one of the two jaws that obtruct
the vacuum chamber after the bending magnet (see fig. \ref{figexp}). This
method allows a resolution of 5\%.
Figure \ref{figspectrum} shows three spectra obtained when only the first cell
is powered at different levels. They all have the same
shape, with two peaks. The lower energy peak corresponds to electrons
directly emitted by the cathode, as shown by PARMELA \cite{bib:parmela}
simulations. The high energy peak corresponds to electrons emitted
by the exit nose of the cavity, accelerated backward to the cathode,
backscattered with most of their impinging energy and fully accelerated
again, thus having almost two times the energy of the electrons
directly emitted by the cathode. This explanation is sustained by simulations
done with the code TW-TRAJ \cite{bib:parodi}.

\fig{Energy spectrum of the dark current emitted by the first cell for three
different peak accelerating fields}
{figspectrum}{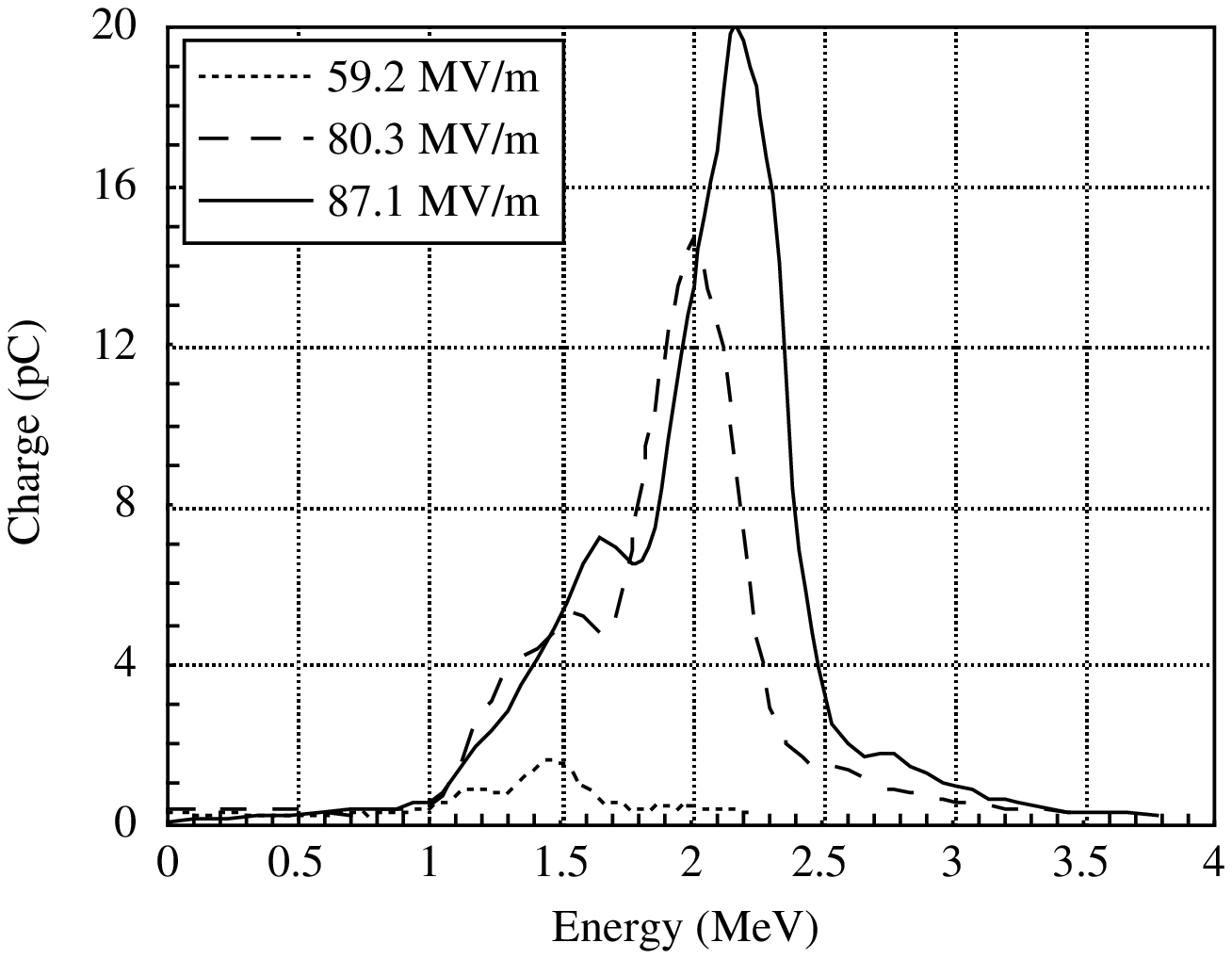}{69.13}{90}{130.41}{425.25}{518.805}{728.595}

\fig{Dark current spectrum from the complete gun for two different phase shifts
between the cells}{figdark}{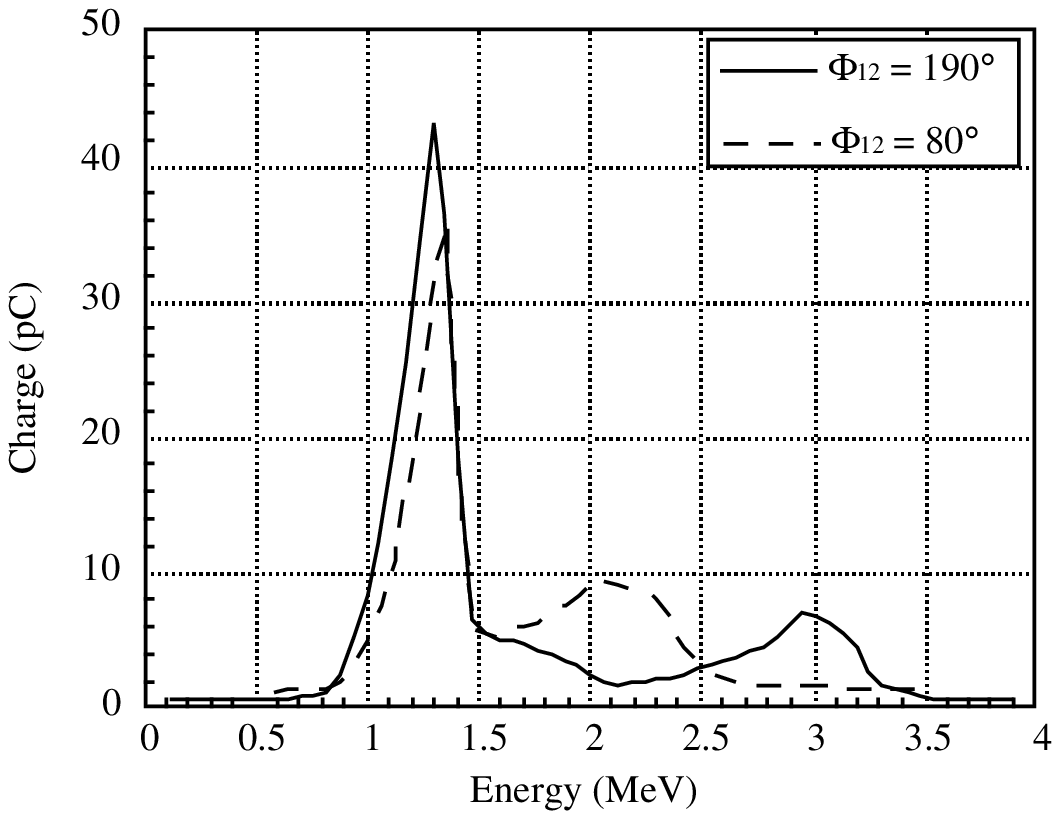}
{69.29}{90}{158.76}{456.435}{479.115}{703.08}

The energy spectrum of the dark current extracted from the complete gun
is more difficult to interpret. Figure \ref{figdark} shows two
spectra obtained for two different phase shifts between the two cells. The low
energy peak
is independent of the phase and therefore is probably due to the second
cell entrance nose. The high energy  peak depends on the phase and thus comes
from electrons emitted in the first cell, either by the cathode or
by the exit nose. The peak centered around 3 MeV probably comes
from electrons emitted by this exit nose and backscattered by the cathode,
since simulations show that cathode electrons can have at maximum an energy of
2.35 MeV.

These energy spectra exhibit somewhat more complicated shapes than
the spectra measured on the Brookhaven RF gun \cite{bib:bnl1}. This is
due to the noses of the cavities and the facility to vary the phase between the
two cavities.

\vspace{0.4cm}
\centerline{\large\bf Photo-electrons measurements}
\vspace{0.4cm}

Since the first photo-electron measurements reported in reference
\cite{bib:lallondres}, very few new measurements could
be made. The only new result is the observation of an intense electron
emission above a certain laser power density.
Figure \ref{figplasma} shows the charge and pulse length observed on the
oscilloscope as a function of the laser energy. In the low laser fluence
regime, the full width  pulse length is 1 ns (limited by cable and oscilloscope
bandwidth) corresponding to the normal photoemission process, with
an efficiency of 1 pC/$\mu$J, as was already measured before
\cite{bib:lallondres}.
Above a certain
energy treshold, the pulse length suddenly increases to 50 ns, and the emitted
charge becomes very high (up to 35 nC). This phenomena has already been
observed at Brookhaven \cite{bib:bnl} where the intense electron bursts
were also measured
to be around 50 ns long, and the laser fluence threshold for their production
was $10^9$ W/cm$^2$.

\fig{Charge and pulse length as a function of laser energy}{figplasma}
{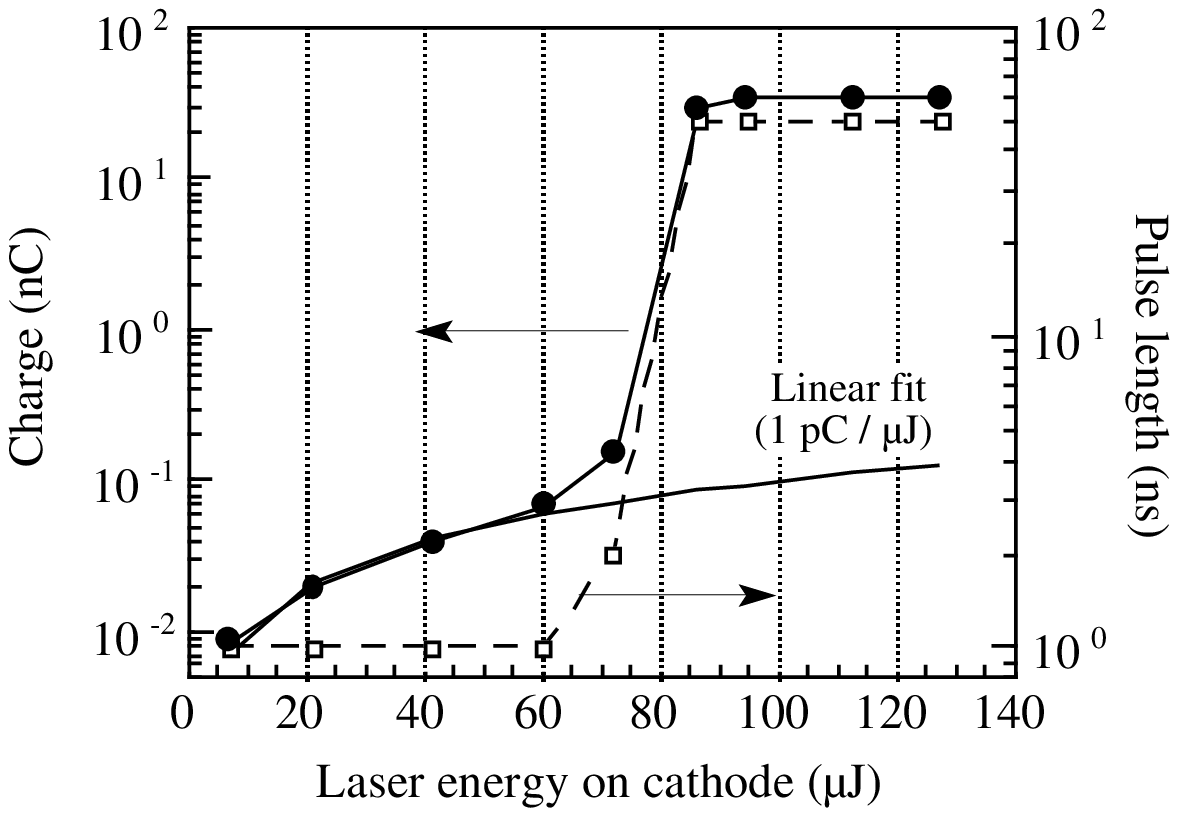}{61.92}{90}{133.245}{459.27}{487.62}{708.75}

In our experiment, we could  not measure precisely the laser spot
size. If we suppose the same fluence threshold as Brookhaven, we can infer
from figure \ref{figplasma}, a laser spot radius of 2 mm, which sounds
reasonable.

By changing the position of the last grating in the laser compressor, it is
possible to vary the laser pulse length in the range 0.5 - 15 ps.
By lengthening the pulse, we checked that it was possible to suppress the
intense emission process. If we  increased the laser spot size, the energy
necessary to obtain the intense emission was changed. These two facts
confirm that the intense emission is related to laser power density.

As we tried to maximize the extracted charge in the normal photoemission
process by enlarging the laser spot size, the largest charge obtained was 240
pC. In practice, this intense emission process limits the use of very short
intense laser pulses with low quantum efficiency cathodes. For example,
with a 2 mm spot diameter, which would  normally be required in CANDELA
gun (knowing that the cavity aperture is only 10 mm), the maximum
charge that one can extract in the photoemission regime is
limited to 16 pC, assuming a quantum efficiency of 5 $\times$ 10$^{-6}$.
Even if one operates the gun at very high gradient on the cathode, thus
taking advantage of the Schottky effect, the charge will still be
limited to less than 200 pC. For this reason, the copper cathode
will soon be replaced by a dispenser cathode that has a measured quantum
efficiency of 5 $\times$ 10$^{-4}$ \cite{bib:leblond}.

\vspace{0.4cm}
\centerline{\large\bf Conclusion}
\vspace{0.4cm}

This paper described two sets of measurements done on CANDELA rf gun:
the dark current spectrum
and the intense emission process obtained for large laser fluence.
This phenomenon not well understood so far, limits the maximum
extractable photemitted charge. It
can be avoided by the use of more efficient cathodes.

\vspace{0.4cm}
\centerline{\large\bf Acknowledgements}
\vspace{0.4cm}

The design and construction of CANDELA involved several people
that we wish to thank. We also thank P. Devlin-Hill for
her participation during the experiments involving the laser, and
T. Garvey for careful reading of this manuscript and interesting
remarks about the intense emission process.

\end{document}